\documentclass[%
 aip,
 amsmath,amssymb,
 reprint,%
]{revtex4-1}

\usepackage{graphicx}
\usepackage{dcolumn}
\usepackage{bm}

\usepackage[utf8]{inputenc}
\usepackage[T1]{fontenc}
\usepackage{mathptmx}
\usepackage{etoolbox}

\usepackage{mylines}
\usepackage{myplots}
\newcommand{\lsy}[3]{\mbox{(\kern-.4em\lineSymbolRGB[#3]{#1}{.8pt}{#2}{4pt}\kern-.4em)}}
\newcommand{\sy}[2]{\mbox{(\kern-.25em\SymbolRGB[solid]{#1}{.8pt}{#2}{4pt}\kern-.25em)}}
\newcommand{\lcap}[2]{~\,{\kern-1em\protect\mylcap{#1}{#2}}}

\definecolor{fd01}{rgb}{0,0,0}
\definecolor{fd04}{rgb}{0.0862,0.3686,0.5172}
\definecolor{fd08}{rgb}{0.0000,0.6980,0.4235}
\definecolor{fd16}{rgb}{1.0000,0.8941,0.0000}
\definecolor{conv}{rgb}{1.0000,0.9098,0.7215}
\definecolor{pool}{rgb}{0.8078,0.3607,0.1372}
\definecolor{prel}{rgb}{0.6627,0.7098,0.3137}
\definecolor{batc}{rgb}{0.4549,0.4470,0.8784}
\definecolor{relu}{rgb}{0.7764,0.8274,0.3058}
\definecolor{subp}{rgb}{0.9294,0.8117,0.8156}

\usepackage[normalem]{ulem}
\usepackage{color}

\makeatletter
\def\@email#1#2{%
 \endgroup
 \patchcmd{\titleblock@produce}
  {\frontmatter@RRAPformat}
  {\frontmatter@RRAPformat{\produce@RRAP{*#1\href{mailto:#2}{#2}}}\frontmatter@RRAPformat}
  {}{}
}%
\makeatother
\begin{document}

\preprint{AIP/123-QED}

\title[From coarse wall measurements to turbulent velocity fields through deep learning]{From coarse wall measurements to turbulent velocity fields\\through deep learning}
\author{A. Güemes}
 \email{guemes.turb@gmail.com}
\affiliation{
Aerospace Engineering Research Group, Universidad Carlos III de Madrid, Leganés, Spain
}%
\author{S. Discetti}%
\author{A. Ianiro}
\affiliation{
Aerospace Engineering Research Group, Universidad Carlos III de Madrid, Leganés, Spain
}%

\author{B. Sirmacek}
\affiliation{%
Smart Cities, School of Creative Technology, Saxion University of Applied Sciences, Enschede, The Netherlands
}%

\author{H. Azizpour}
\affiliation{%
Division of Robotics, Perception, and Learning, KTH Royal Institute of Technology, Stockholm, Sweden
}%
\affiliation{%
Swedish e-Science Research Centre (SeRC), Stockholm, Sweden
}%

\author{R. Vinuesa}
\affiliation{%
SimEx/FLOW, Engineering Mechanics, KTH Royal Institute of Technology, Stockholm, Sweden
}%
\affiliation{%
Swedish e-Science Research Centre (SeRC), Stockholm, Sweden
}%

\date{\today}

\begin{abstract}
This work evaluates the applicability of super-resolution generative adversarial networks (SRGANs) as a methodology for the reconstruction of turbulent-flow quantities from coarse wall measurements.
The method is applied both for the resolution enhancement of wall fields and the estimation of wall-parallel velocity fields from coarse wall measurements of shear stress and pressure.
The analysis has been carried out with a database of a turbulent open-channel flow with friction Reynolds number $Re_{\tau}=180$ generated through direct numerical simulation. 
Coarse wall measurements have been generated with three different downsampling factors $f_d=[4,8,16]$ from the high-resolution fields, and wall-parallel velocity fields have been reconstructed at four inner-scaled wall-normal distances $y^+=[15,30,50,100]$. 
We first show that SRGAN can be used to enhance the resolution of coarse wall measurements.
If compared with direct reconstruction from the sole coarse wall measurements, SRGAN provides better instantaneous reconstructions, both in terms of mean-squared error and spectral-fractional error.
Even though lower resolutions in the input wall data make it more challenging to achieve highly accurate predictions, the proposed SRGAN-based network yields very good reconstruction results.
Furthermore, it is shown that even for the most challenging cases the SRGAN is capable of capturing the large-scale structures that populate the flow.
The proposed novel methodology has great potential for closed-loop control applications relying on non-intrusive sensing. 
\end{abstract}

\maketitle

\section{Introduction}
In recent years, research in deep neural networks (DNNs) has been fueled by new available computational resources, which have brought a wide variety of new techniques for visual object recognition, object detection and speech recognition among many others~\cite{lecun2015deep}.
The rise of DNNs in many applications, such as medicine \cite{zeleznik2021deep,de2018clinically}, climate \cite{waldmann2019mapping}, wildlife ecology \cite{norouzzadeh2018automatically}, physics \cite{udrescu2020ai,kwon2020magnetic} or sustainability \cite{vinuesa_et_al_2020}, has not been overlooked in fluid-mechanics research \cite{kutz2017deep,brunton2020machine}. 
Some of the outstanding applications of DNNs in fluid mechanics are the improvement of Reynolds-averaged Navier--Stokes simulations \cite{ling2016reynolds}, the extraction of turbulence theory for two-dimensional flow \cite{jimenez2018machine}, prediction of temporal dynamics~\cite{srinivasan2019predictions,eivazi2021recurrent} or the embedding of physical laws in DNN predictions \cite{raissi_et_al}. 

Generative adversarial networks (GANs), firstly introduced in Ref.~\onlinecite{goodfellow2014generative}, are one of the latest advances in DNN research. 
Based on game theory, GANs are composed of two competing networks: a generator that tries to produce an artificial output which mimics reality; and a discriminator, which is in charge of distinguishing between reality and artificial outputs. 
During training, the generator network makes its output more realistic by improving the features that the discriminator identified as artificial.
Among the different areas in which GANs have been applied successfully, their use to enhance image resolution stands out \cite{ledig2017photo,wang2018esrgan}. 
In fluid-mechanics research, they have been successfully applied to recover high-resolution fields in different types of flow, such as the wake behind one or two side-by-side cylinders \cite{deng2019super} or volumetric smoke data \cite{werhahn2019multi}. 
While in these works the training has been carried out with a supervised approach i.e., with paired high- and low-resolution flow fields, GANs have been recently applied with an unsupervised approach to enhance the resolution of homogeneous turbulence and channel flows \cite{kim2021unsupervised}.
GANs are now challenging other resolution-enhancement strategies based on convolutional neural networks (CNNs), which showed to be successful for the cases of the flow around a cylinder, two-dimensional decaying isotropic turbulence \cite{fukami2019super} and channel flows \cite{liu2020deep}.
More recently, Ref.~\onlinecite{fukami2021machine} has proposed a methodology based on CNNs to recover high-resolution sequences of flow fields in homogeneous isotropic and wall turbulence from the low-resolution fields at the beginning and end of the sequence.

CNNs have also been used successfully to estimate flow fields using field measurements of wall shear and/or pressure. 
Several methods have been proposed, such as the direct reconstruction of the flow field from the wall quantities using fully-convolutional networks (FCNs) ~\cite{guastoni,guastoni2020convolutional}, or the use of proper orthogonal decomposition\cite{lumley1967structure} (POD) in combination with CNNs \cite{guemes2019sensing} and FCNs \cite{guastoni2020convolutional}. 
Moreover, Ref.~\onlinecite{guemes2019sensing} studied the effect of the wall-resolution measurements on the predictions accuracy, showing that their architecture was able to continue providing predictions of similar accuracy for downsampling factors up to 4. 
When a limited number of sensors is available, shallow neural networks (SNNs) offer another option for this task. 
Ref.~\onlinecite{erichson2020shallow} compared SNNs with POD for the reconstruction of a circular cylinder wake, sea surface temperature, and decaying homogeneous isotropic turbulence, showing that the new data-driven approach outperforms the traditional one.

In the first part of the present work, a GAN-based methodology is proposed to recover high-resolution fields of wall measurements. 
Because the results are very positive when performing this task, and it has already been shown that GANs can be used successfully in enhancing turbulent-flow resolution\cite{kim2021unsupervised}, the second part of this work extends their use to reconstruct high-resolution wall-parallel flow fields from coarse wall measurements.
This method is compared with the FCN-POD architecture proposed in Ref.~\onlinecite{guastoni2020convolutional}.
The choice is based on the proven capability of this network to deal with low-resolution input information \cite{guemes2019sensing}.
The paper is organized as follows: \S\ref{sec:metho} outlines the details of the numerical database used for this study and presents the different DNNs employed for that purpose; the main results for wall-resolution enhancement are provided in \S\ref{sec:wall}, while the flow-reconstruction results are reported in \S\ref{sec:flow}.
To close the paper, \S\ref{sec:concl} presents the main conclusions of the work.

\section{Methodology}\label{sec:metho}

This section presents the details of the numerical database employed for this study, as well as the DNN architectures and the training methodology with which they have been optimized. 
Throughtout the paper $x$, $y$, and $z$ denote the streamwise, wall-normal, and spanwise directions respectively, with $u$, $v$, and $w$ referring to their corresponding instantaneous velocity fluctuations. 
Streamwise and spanwise wall-shear-stress fluctuations are referred to as $\tau_{w_x}$ and $\tau_{w_z}$ respectively, with $p_w$ denoting the pressure fluctuations at the wall.

\subsection{Dataset description}

The methodology proposed in this work has been tested with a direct numerical simulation (DNS) of a turbulent open-channel flow generated with the pseudo-spectral code SIMSON \citep{chevalier}. 
The simulation domain extends $4\pi h \times h \times 2\pi h$ (where $h$ is the channel height) in the streamwise, wall-normal and spanwise directions respectively, with the flow represented by 65 Chebyshev modes in the wall-normal direction and with 192 Fourier modes in the streamwise and spanwise directions. 
The simulation is characterized by a friction Reynolds number $Re_{\tau}=180$, which is based on $h$ and the friction velocity $u_{\tau}=\sqrt{\tau_w/\rho}$ (where $\tau_w$ is the magnitude of the wall-shear stress and $\rho$ is the fluid density).
The superscript '+' denotes inner-scaled quantities, using $u_{\tau}$ for the velocities and the viscous length $\ell^*=\nu/u_{\tau}$ (where $\nu$ is the fluid kinematic viscosity) for the distances.
DNNs have been trained with 50,400 samples separated by $\Delta t^+=5.08$, while 3,125 samples with time separation $\Delta t^+=1.69$ have been used for testing them.
For further simulation details, see Ref.~\onlinecite{guastoni2020convolutional}.

Wall information, used as input to reconstruct wall-parallel fluctuating velocity fields, is composed of streamwise and spanwise shear stress, as well as pressure fluctuations. 
To assess the capability of our methodology to reconstruct turbulent wall and flow fields from coarse wall measurements, three different sets of downsampled wall fields have been generated, with downsampling factors $f_d=[4,8,16]$. 
Note that $f_d$ is defined as the resolution reduction in each direction, thus a downsampling factor of $f_d$ yields a downsampled field with a number of points equal to $f_d^{-2}$ times the original.
It has to be noted that $f_d$ values of 2 and 4 were considered in Ref.~\onlinecite{guemes2019sensing}, although for a test case with larger $Re_{\tau}$. 
The reconstruction of the fluctuating velocity fields is evaluated at four different inner-scaled wall-normal distances: $y^+=[15,30,50,100]$. 

\subsection{Super-resolution generative adversarial networks}
\begin{figure*}
  \centerline{
  \includegraphics[width=\linewidth]{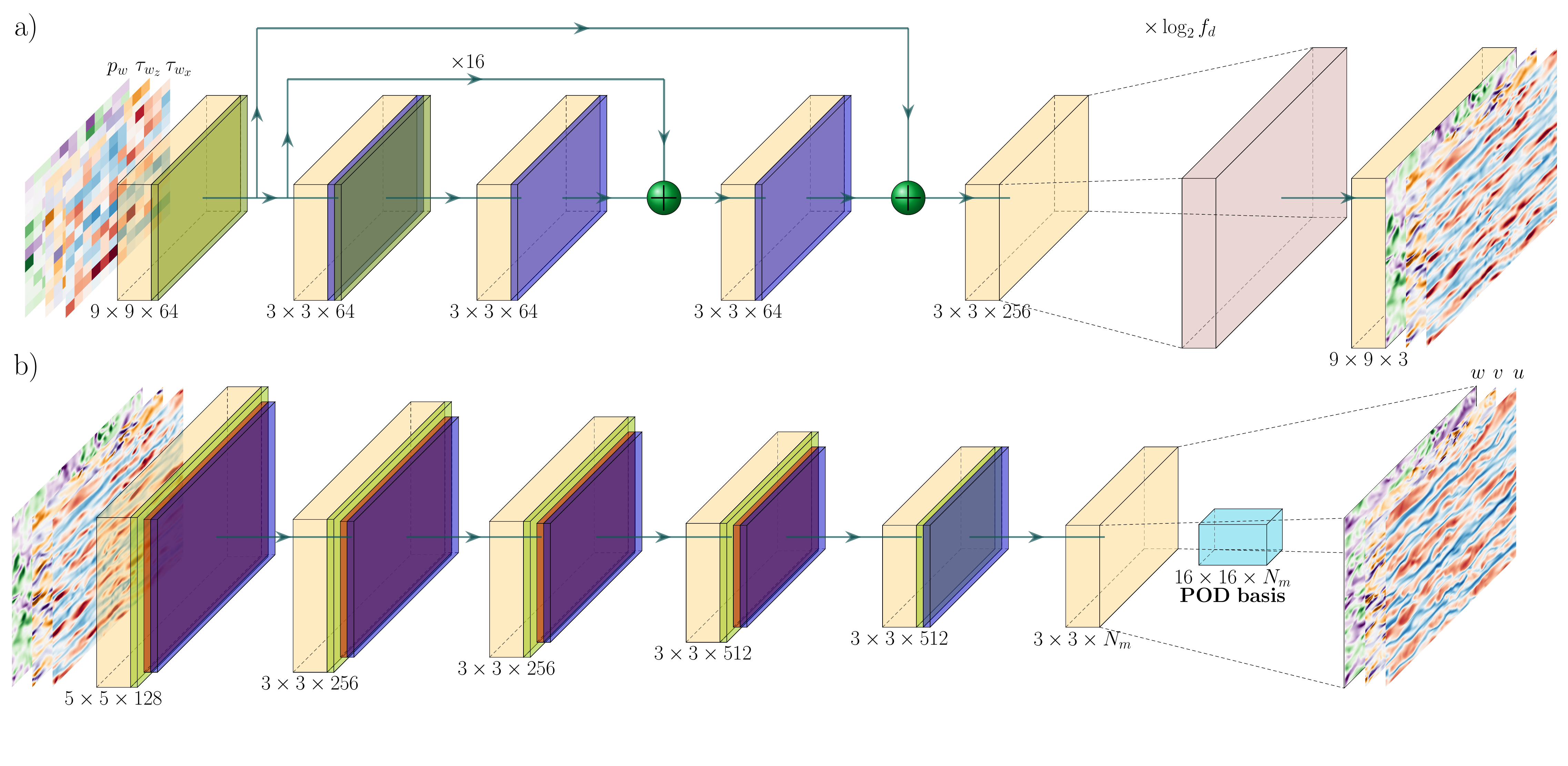}}
  \caption{Schematic view of the DNN architectures for a) generator network in SRGAN, and b) FCN-POD. The colour coding for each layer is: 2D-convolution \sy{conv}{b}, parametric-ReLU-activation \sy{prel}{b}, batch-normalization \sy{batc}{b}, sub-pix-convolution \sy{subp}{b}, ReLU-activation \sy{relu}{b}, and max-pooling \sy{pool}{b} layers. The kernel size and the number of filters are shown at the bottom of the convolution layers.}
\label{fig:01}
\end{figure*}

Super-resolution GAN (SRGAN) is proposed as a method to reconstruct turbulent wall-measurement fields.
Additionally, SRGANs are explored also for direct estimation of velocity fields in wall-parallel planes.
A typical SRGAN architecture consists of two networks: a generator ($G$) and a discriminator ($D$); $G$ is in charge of generating a high-resolution artificial image $\widetilde{H}_R$ from its low-resolution counterpart $L_R$, whereas $D$ is in charge of distinguishing between high-resolution real images $H_R$ and artificial ones. 
Note that the purpose of this work is not to generate a custom architecture to tackle fluid-mechanics cases, since these types of DNNs are already available in the literature~\cite{deng2019super,werhahn2019multi,kim2021unsupervised}. 
Therefore, the architecture presented in Ref.~\onlinecite{ledig2017photo} was used in this study.
It uses a CNN as generator, where the main core is composed of 16 residual blocks \cite{he2016deep}, and the resolution increase is carried out at the end of the network by means of $\log_2(f_d)$ sub-pixel convolution layers \citep{shi2016real}. 
In the case of flow-field reconstruction with full-resolution wall data as input, the sub-pixel convolution layers are removed.
For the discriminator, convolution layers are also used before adding two fully-connected layers, using a sigmoid activation in the last one to obtain a probability to discern whether the high-resolution input is real or not. 
A schematic view of the generator network is shown in Figure~\ref{fig:01}a) and the rest of details can be found in Ref.~\onlinecite{ledig2017photo}. 
The discriminator loss is defined as:
\begin{equation}
    \mathcal{L}_D=-\mathbb{E}[\log D(H_R)] - \mathbb{E}[\log(1-D(G(L_R)))].
\end{equation}

For the generator loss, we have used the perceptual loss \cite{ledig2017photo}, where the content loss is evaluated with the pixel-based mean-squared error between $H_R$ and $\widetilde{H}_{R}$, leading to:
\begin{equation}
    \mathcal{L}_G=\frac{1}{N_xN_z}\sum^{Nx}_{i=1}\sum^{Nz}_{j=1}|G(L_R)_{i,j} - H_R{_{i,j}}|^2 - \lambda \mathcal{L}_D,
\end{equation}
\noindent where $N_x$ and $N_z$ are the number of grid points in the streamwise and spanwise directions for the high-resolution images (192 for both of them in our case) and $\lambda$ is a scalar to weight the value of the adversarial loss, set to $10^{-3}$. 
The weights of the model for each downsampling case have been optimized for 30 epochs using the Adam algorithm~\cite{kingmaba} with learning rate $10^{-4}$.

\subsection{POD-based fully-convolutional networks}
The baseline method for assessing the quality of flow reconstructions achieved by SRGAN is the FCN-POD approach \cite{guemes2019sensing,guastoni2020convolutional}. 
This method divides the turbulent flow fields into $N_s$ two-dimensional subdomains of $N_p\times N_p$ grid points, and POD is performed on each of these subdomains.
The number of subdomains is chosen based on $Re_{\tau}$, with the purpose of ensuring that ~90\% of the flow kinetic energy is contained within $\mathcal{O}(10^2)$ POD modes which can be represented by convolutional filters. 
For the  $Re_{\tau}=180$ case, each field is divided into $12\times12$ subdomains, each of them with $16\times16$ grid points.
The proposed architecture will reconstruct this three-dimensional tensor of POD coefficients from the wall quantities; this tensor is later converted into the flow field by projecting the POD coefficients of each subdomain into its corresponding basis. 
Note that this method does not ensure continuity between subdomains; nevertheless, the convolutional layers have been shown to provide reasonably smooth flow fields\cite{guastoni2020convolutional}.
For each wall-normal distance a different model has been used, the weights of which have been optimized for 30 epochs using the Adam optimizer \cite{kingmaba} with $\epsilon=0.1$, learning rate $10^{-3}$ and an exponential decay starting from epoch 10. 
A schematic representation of the architecture is shown in Figure \ref{fig:01}b), and the rest of the implementation details can be found in Ref.~\onlinecite{guastoni2020convolutional}. 
For the case of coarse input data, a modified version of the FCN-POD model has been used. 
To deal with the different sizes of the input and output tensors, $\log_2(f_d)$ pooling layers have been removed from the original model. 

\section{Assessment of resolution enhancement for wall measurements}\label{sec:wall}

\begin{figure*}
  \begin{center}
  \includegraphics[width=\linewidth]{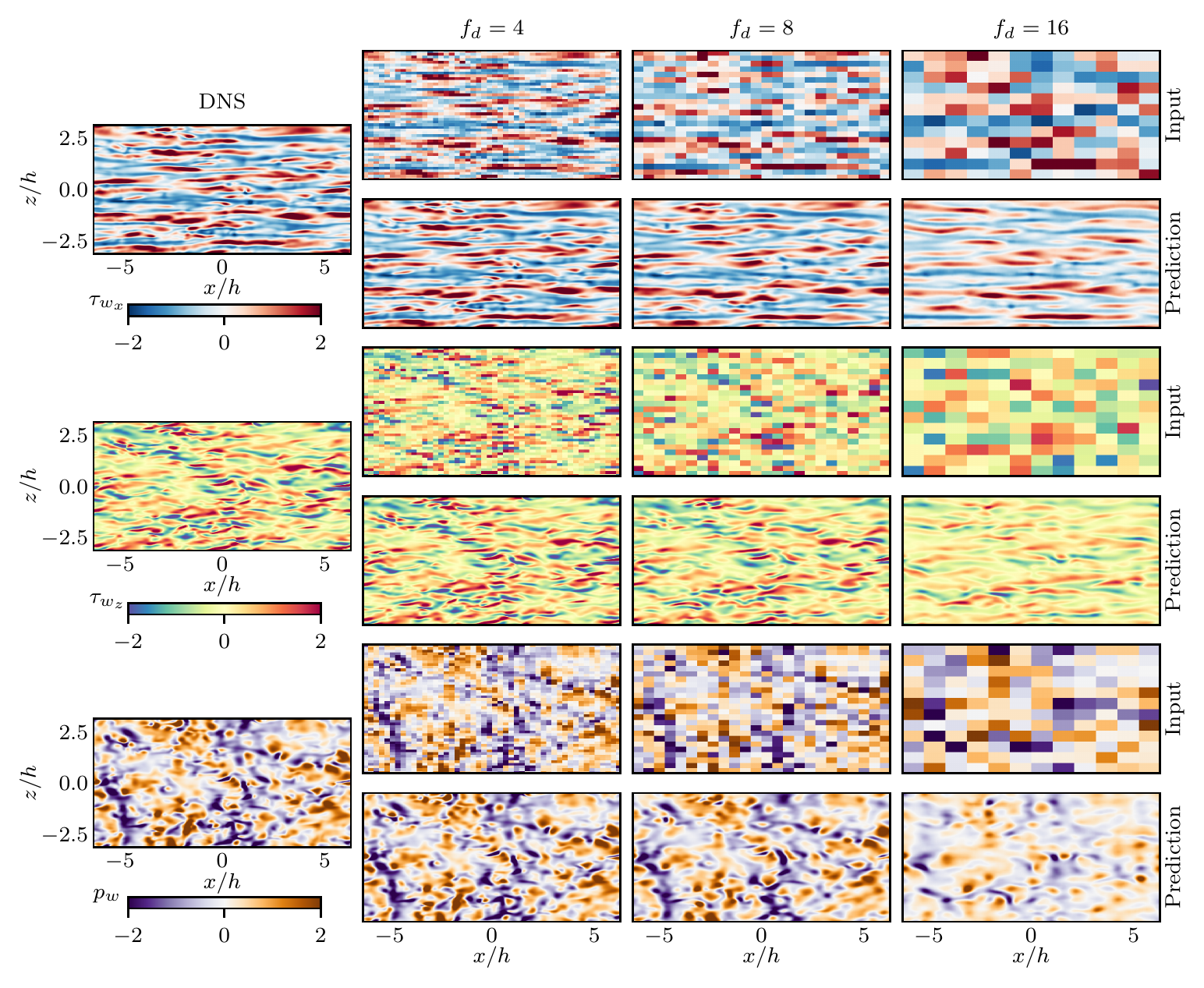}
  \end{center}
  \caption{Comparison of the wall-quantity fluctuating fields at $Re_{\tau} = 180$, scaled with their corresponding standard deviation. Reference DNS is reported at left panel, while the six-row panels report the different $f_d$ cases, covering $f_d=4$ (left), $f_d=8$ (center), and $f_d=16$ (right). Odd rows refer to low-resolution inputs, and even ones to the SRGAN predictions. Top two-row panels report streamwise wall-shear stress, middle ones report spanwise wall-shear stress and bottom ones refer to pressure fluctuations.}
\label{fig:02}
\end{figure*}

The quality of the resolution enhancement of the wall fields is evaluated first. 
Figure~\ref{fig:02} shows an instantaneous field of the streamwise and spanwise wall-shear-stress  and pressure fluctuations for the DNS reference and the SRGAN predictions. 
While the reconstructions from fields with $f_d=4$ and $f_d=8$ recover almost all the flow features present in the DNS references, the instantaneous field for $f_d=16$ exhibits loss of small-scale details. 
Moreover, it appears that the high-intensity regions are attenuated for the latter case. 
Note however, that the locations and sizes of the largest flow structures are very well represented even for $f_d=16$. 

The first observations on the resolution-enhancement performance with respect to $f_d$ obtained from the inspection of instantaneous fields are confirmed when analyzing the mean-squared-error of those fields. 
The errors, normalized with the standard deviation of each quantity, are reported in Table~\ref{tab:01}. SRGANs show excellent results for $f_d=4$ in the three wall quantities, and confirm the performance decay between $f_d=8$ and $f_d=16$.
When assessing the performance differences among wall quantities, it is clear that with larger downsampling factors the errors in the streamwise wall-shear-stress fields are lower than for the other two wall quantities.
This behaviour can be ascribed to the spatial organization of streawise  wall-shear-stress fluctuations, which exhibit a characteristic alignment in the streamwise direction.

\begin{table}
\caption{\label{tab:01}Mean-squared-error in the instantaneous wall fields scaled with their corresponding standard deviations.}
\begin{ruledtabular}
\begin{tabular}{lccc}
$f_d$ & $\tau_{w_x}$ & $\tau_{w_z}$ & $p_w$\\
\hline
4     & 0.0187       & 0.0244       & 0.0153 \\
8     & 0.2240       & 0.3041       & 0.2741 \\
16    & 0.6531       & 0.7732       & 0.7461 \\ 
\end{tabular}
\end{ruledtabular}
\end{table}

\begin{figure}
  \centerline{\includegraphics[width=\columnwidth]{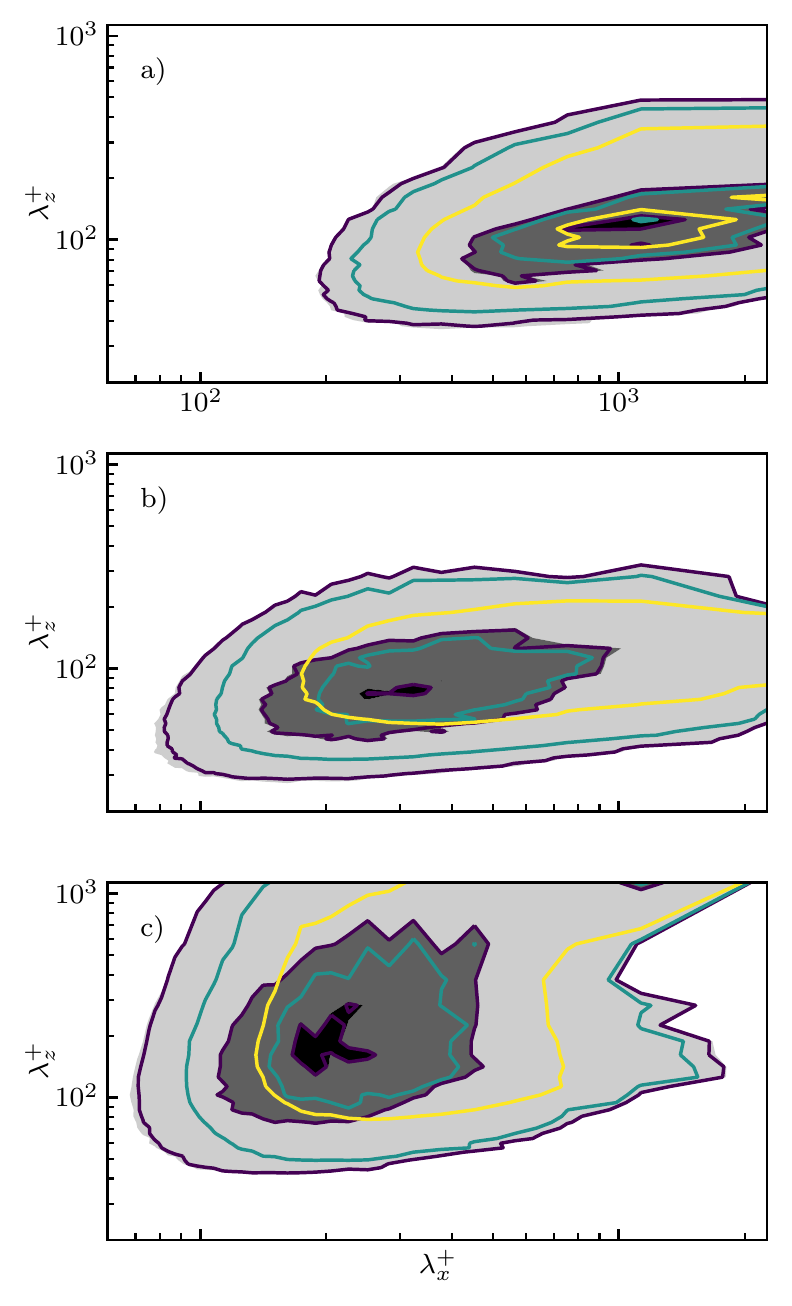}}
  \caption{Pre-multiplied two-dimensional power-spectral densities for a) fluctuating streamwise wall-shear-stress, b) fluctuating spanwise wall-shear-stress, and c) wall-pressure fluctuations. The contour levels contain 10\%, 50\% and 90\% of the maximum DNS power-spectral density. Shaded contours refer to the reference DNS data, while coloured lines denote $f_d=4$ \lcap{-}{fd04}, $f_d=8$ \lcap{-}{fd08}, and $f_d=16$ \lcap{-}{fd16}.}
\label{fig:03}
\end{figure}

The pre-multiplied two-dimensional inner-scaled spectra for the three wall quantities are reported in Figure~\ref{fig:03}. 
The high-energy peak containing 90\% or more of $\tau_{w_x}$ is well captured by the predictions with $f_d=4$ and $f_d=8$, while for $f_d=16$ this is not recovered, even showing a significant attenuation of 50\% of the total energy content. 
The energy attenuation is even stronger for  $\tau_{w_z}$ and $p_w$, where the predictions of $f_d=4$ are the only ones capturing the energy distribution for both quantities. 
In the case of $f_d=16$, the attenuation is so significant that even the 50\% energy-content level is not recovered.
The distribution of scales over a larger range also explains why the $\tau_{w_x}$ error is smaller than that of the other two wall quantities, since for the first one the SRGAN architecture deals with a lower parametric space.

Although the scope of this work is not to develop a customized SRGAN architecture for wall turbulence, here we briefly compare our results with those of other studies in the literature.
For example, Ref.~\onlinecite{kim2021unsupervised} used an unsupervised GAN to reconstruct wall-parallel velocity fields at $y^+=15$ and $y^+=100$ in a turbulent channel flow at $Re_{\tau}=1000$ with $f_d=8$. 
They report good resolution-enhancing results in terms of instantaneous fields, turbulence statistics and one-dimensional spectra, similar to ours for the same $f_d$. 
With respect to the spectra, their work and ours coincide in identifying the small-scale structures as those most difficult to recover. 
Because of the different $Re_{\tau}$ in both studies, it is important to highlight that $f_d$ is a pixel ratio between the high- and low-resolution fields, and it does not take into account how many viscous lengths are contained in a single pixel. 
For a fair comparison in turbulent flows, we propose the following normalized downsampling factor:
\begin{equation}
    \tilde{f}_{d}=f_d\sqrt{\Delta x^{{+}^2} + \Delta z^{{+}^2}},
    \label{eq:03}
\end{equation}

\noindent where $\Delta x^{+}$ and $\Delta z^{+}$ are the inner-scaled grid spacing in physical space for $x$ and $z$ respectively.
Using equation~(\ref{eq:03}) yields a normalized downsampling factor $\tilde{f}_{d}\approx 105$, while the work of Ref.~\onlinecite{kim2021unsupervised} tackles a problem with $\tilde{f}_{d}\approx 109$, therefore showing that the comparison is fair.

\section{Prediction of turbulent flow fields \\from coarse wall measurements}\label{sec:flow}

This section presents the reconstruction performance of wall-parallel velocity fields from wall measurements.
Reconstructions are performed at four different wall-normal distances: $y^+=[15,30,50,100]$, considering 4 downsampling cases for the input wall measurements: $f_d=[1,4,8,16]$.
Note that $f_d=1$ means that no information is lost at the wall with respect to the DNS reference.

\begin{figure*}
  \centerline{\includegraphics[width=\textwidth]{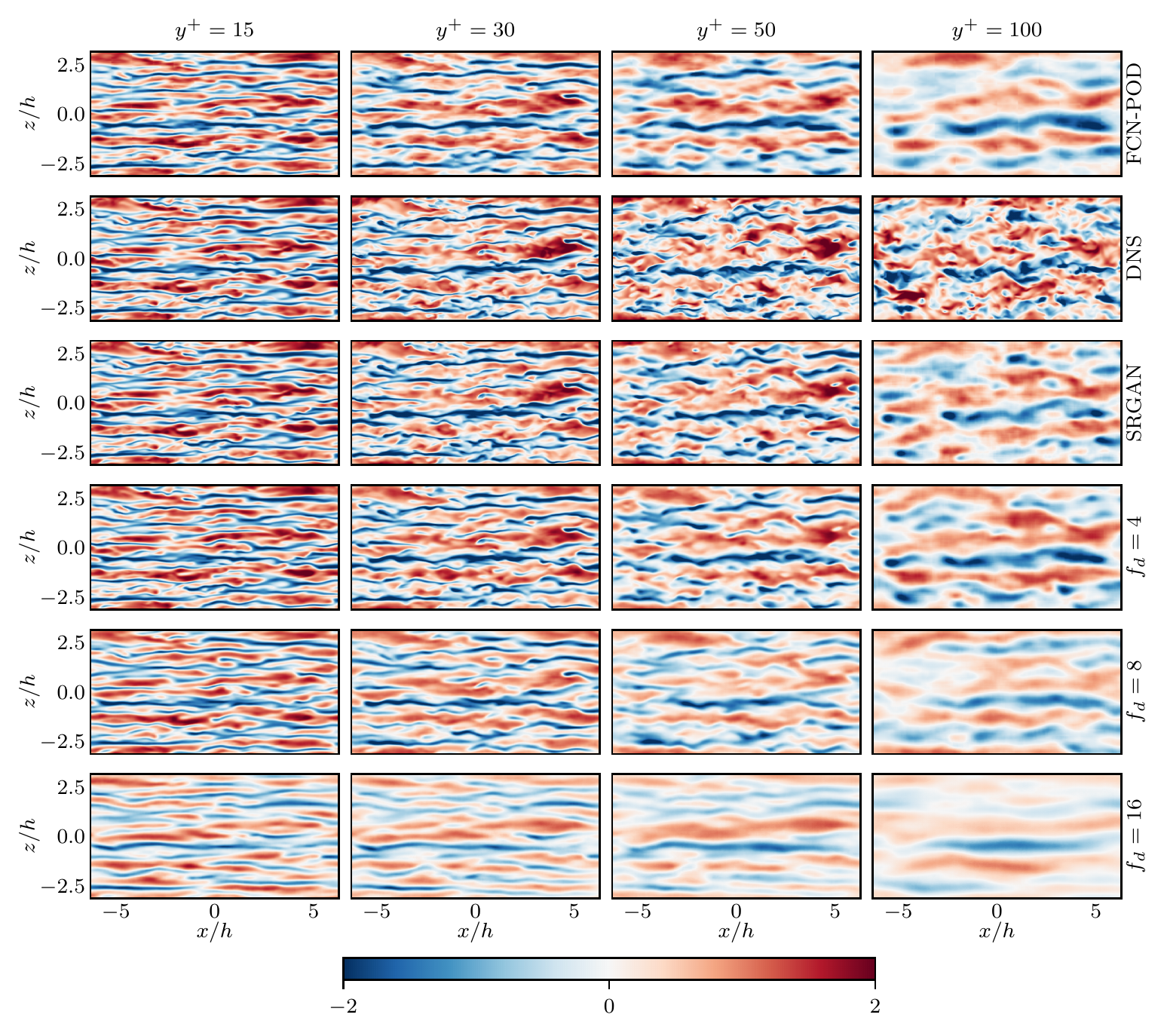}}
  \caption{Countour map for the streamwise velocity fluctuation fields scaled with the corresponding standard deviation. From top to bottom, rows denote FCN-POD, reference DNS, and SRGAN predictions with $f_d=[1,4,8,16]$ for the wall information. From left to right, columns indicate $y^+=15$, $y^+=30$, $y^+=50$ and $y^+=100$.}
\label{fig:04}
\end{figure*}

Figure~\ref{fig:04} shows instantaneous fields of the streamwise velocity fluctuations at the four wall-normal distances of interest in this study.
Predictions generated with SRGAN and FCN-POD networks are compared to the DNS reference.
Note that the FCN-POD predictions are only reported for $f_d=1$ case, and they are analogous to the results presented in Ref.~\onlinecite{guastoni2020convolutional}.
Inspecting the fields, it can be seen that the best results are obtained closer to the wall, with the lowest downsamplings. 
When moving away from the wall or reducing the information provided by the wall, the small-scale fluctuations in the fields start to disappear, and one of the networks recover the high-intensity fluctuating regions of the flow. 
However, there is a clear performance difference between FCN-POD and SRGAN predictions.
While the loss of small-scale fluctuations is clearly observable at $y^+=30$ for the FCN-POD predictions, SRGAN is able to capture most of them at $y^+=50$.
It is not until $y^+=100$ that it is clearly observed that small-scale fluctuations are not recovered by the SRGAN architecture.
Nonetheless, the results of $f_d=8$ and $f_d=16$ at $y^+=15$ are successful in capturing most of the flow features present in the DNS reference, and the same can be said for $f_d=8$ at $y^+=30$. 
Since most of the flow-control techniques actuate over this region \cite{choi1994active,lee1997application,bai2014active}, these results indicate that equally-distributed probes would be sufficient to feed flow information to these control techniques, instead of using image-based acquisition systems, which are more expensive and difficult to implement. 

\begin{figure}
  \centerline{\includegraphics[width=\columnwidth]{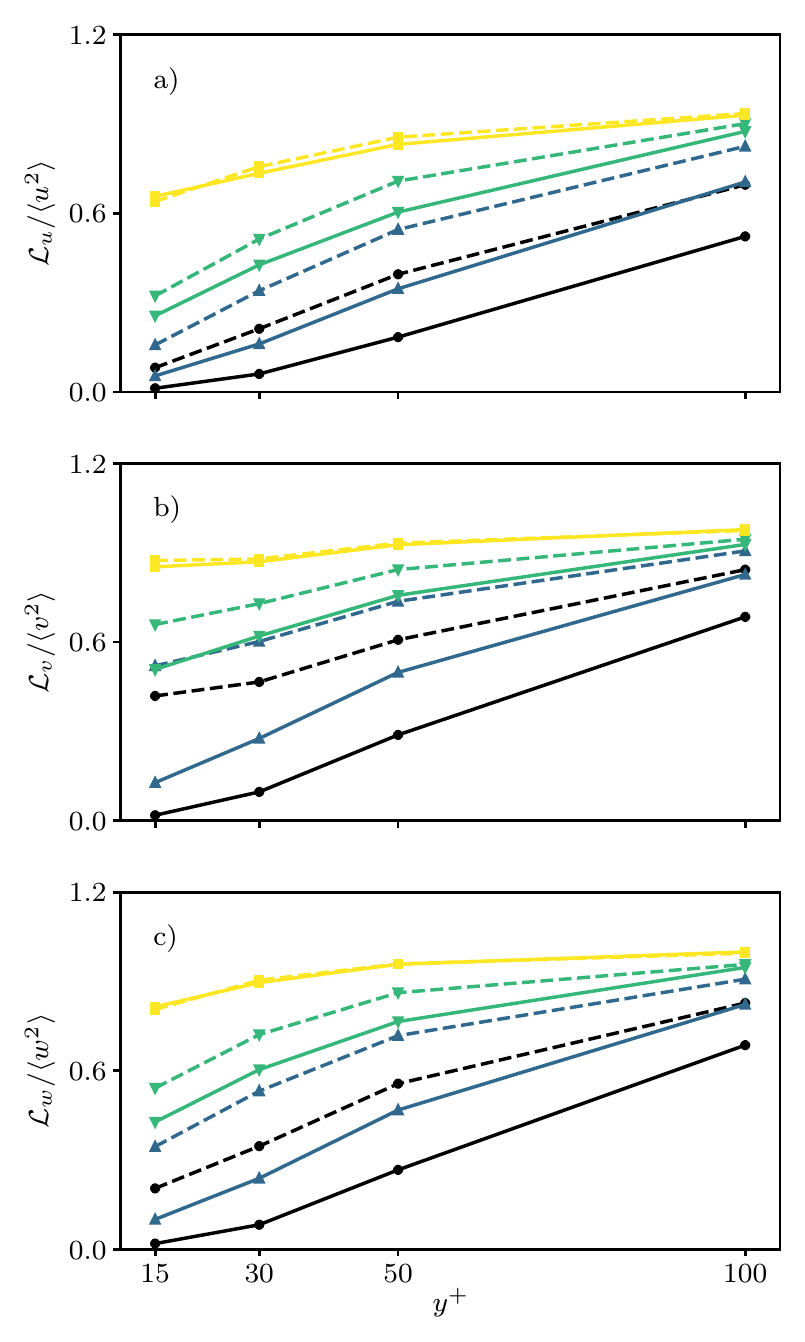}}
  \caption{Mean-squared-error in the instantaneous fields of a) streamwise, b) wall-normal, and c) spanwise velocity fluctuations scaled with their corresponding standard deviations. Line styles refer to \lcap{-}{fd01} SRGAN, and \lcap{--}{fd01} FCN-POD predictions, respectively. Colours and symbols denote $f_d=1$ \sy{fd01}{o*}, $f_d=4$ \sy{fd04}{t*}, $f_d=8$ \sy{fd08}{dt*}, and $f_d=16$ \sy{fd16}{s*}.}
\label{fig:05}
\end{figure}

A global view of the flow-reconstruction performance is provided in terms of mean-squared-error. 
Figure~\ref{fig:05} reports the evolution of the error with respect to the wall-normal distance for the three flow quantities, the four $f_d$ values and the two reconstruction techniques. 
There are two aspects to analyze: the performance difference between the two networks, and the evolution of the error with respect to $f_d$. 
When comparing the error evolution for both networks, it can be seen that SRGAN outperforms FCN-POD predictions for all $f_d=4$ and $f_d=8$ cases, where the errors for the predictions generated with SRGAN are better than when using FCN-POD approach. 
However, for the $f_d=16$ case both errors collapse, and therefore the benefit of using SRGAN disappears. 
This deterioration of the flow reconstruction can be ascribed to the low amount of information contained by the coarse wall measurements. 
In the case of $f_d=1$, the wall data contains information of the fluctuations with characteristic lengths as small as $\sim10\ell^*$ in the streamwise direction, while for $f_d=16$ this increases up to $\sim160\ell^*$.
While $f_d=4$ and $f_d=8$ recover the small scales present in the DNS reference, $f_d=16$ does not succeed in this task.
In any case, it is important to remark the significant accuracy improvement of SRGAN for $f_d=1$ with respect to the FCN-POD method, especially in the wall-normal and spanwise components.
Although not presented here, this improvement is noticeable also if compared with the FCN method used in Ref.~\onlinecite{guastoni2020convolutional}.

\begin{figure*}
  \centerline{\includegraphics[width=\textwidth]{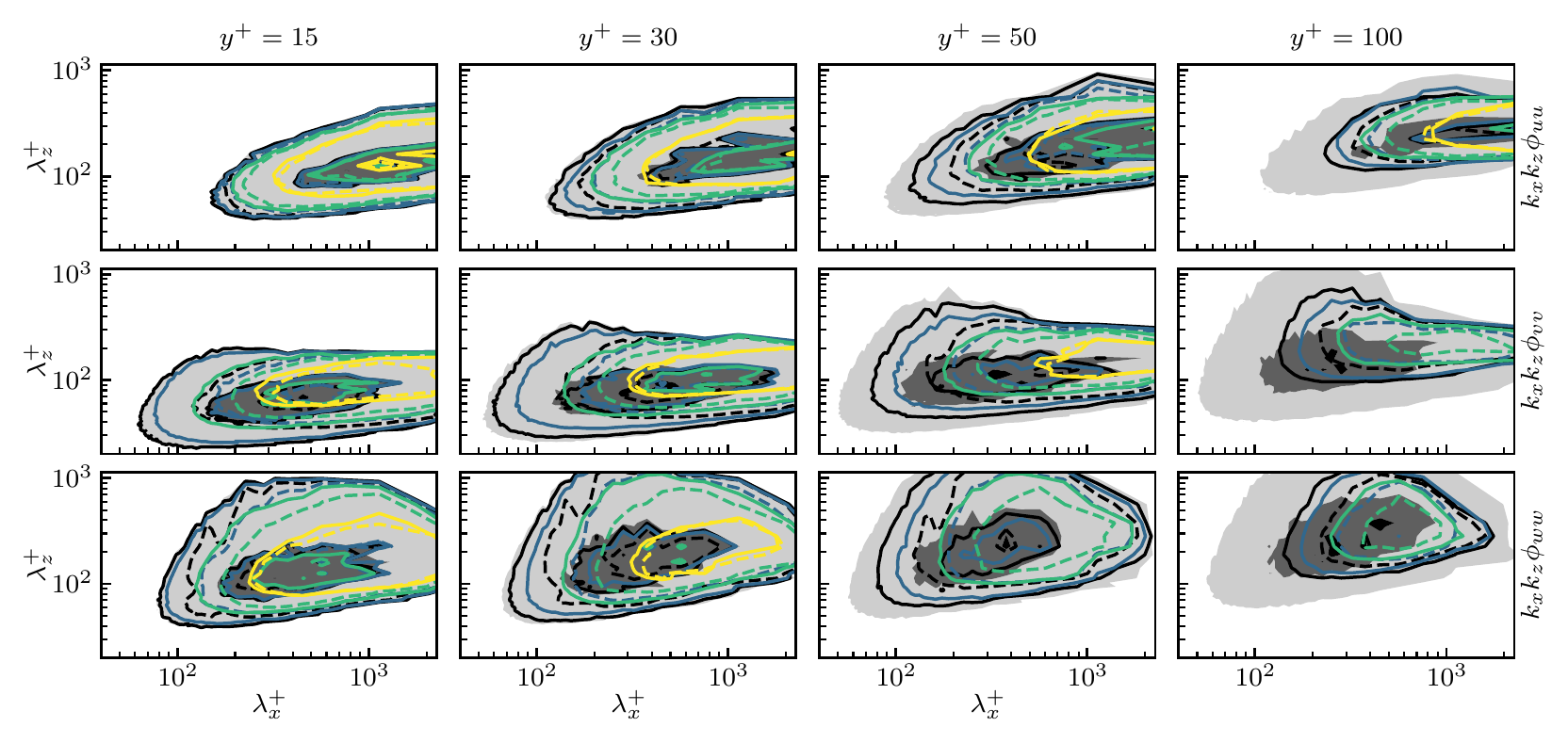}}
  \caption{Pre-multiplied two-dimensional power-spectral densities for streamwise (first row), wall-normal (second row), and spanwise (third row) velocity fluctuations. From left to right, columns refer to inner-scaled wall distance $y^+$ equal to 15, 30, 50, and 100. The contour levels contain 10\%, 50\% and 90\% of the maximum DNS power-spectral density. Shaded contours refer to the reference DNS data, while contour lines refer to \lcap{-}{fd01} SRGAN-FCN-POD, and \lcap{--}{fd01} FCN-POD predictions, respectively. Colours denote $f_d=1$ \sy{fd01}{s*}, $f_d=4$ \sy{fd04}{s*}, $f_d=8$ \sy{fd08}{s*}, and $f_d=16$ \sy{fd16}{s*}.}
\label{fig:06}
\end{figure*}

The second factor to analyze is the performance decay of the predictions when increasing $f_d$. 
In a previous study\cite{guemes2019sensing}, the effect of $f_d$ when reconstructing the large-scale structures present in wall-parallel flow fields from wall measurements on a turbulent channel flow of $Re_{\tau}=1000$ was analyzed. 
The analyzed effect of $f_d=[1,2,4]$ reported only a weak deterioration effect due to the increase of $f_d$. 
However, the results presented in Figure~\ref{fig:06} show a clear dependency between $f_d$ and the mean-squared-error. Once again, the question arises whether $f_d$ is adequate to characterize the downsampling effect in wall turbulence. 
If we used the normalized downsampling factor proposed in equation~(\ref{eq:03}), $f_d=4$ becomes $\tilde{f}_{d}\approx44$ for Ref.~\onlinecite{guemes2019sensing}, while in our case it is $\tilde{f}_{d}\approx52$, increasing to 105 and 210 for $f_d$ values equal to 8 and 16 respectively. 
Therefore, it can be argued that in this work we are facing a more challenging wall-information loss. 
Furthermore, it must be recalled that the flow scales to be predicted also affect the performance of the method.
Ref.~\onlinecite{guemes2019sensing} only targeted the flow scales in the first 10 POD modes, while this work targets the entire energy spectra.
The first 10 POD modes of Ref.~\onlinecite{guemes2019sensing} refer to the most energetic structures present in the flow. 
Large coherent structures are more persistent over time, with lives proportional to their scale\cite{lozano2014time}. 
These characteristic length and time scales make them less sensitive to the changes in the resolution of the wall data. 
However, small-scale structures are affected, both because they are smaller than the scales contained in the coarse wall data and because the modulation effect of large scales~\cite{hutchins2007large,dogan_modulation} is also hidden by the low-resolution data.

\begin{figure*}[!t]
  \centerline{\includegraphics[width=\textwidth]{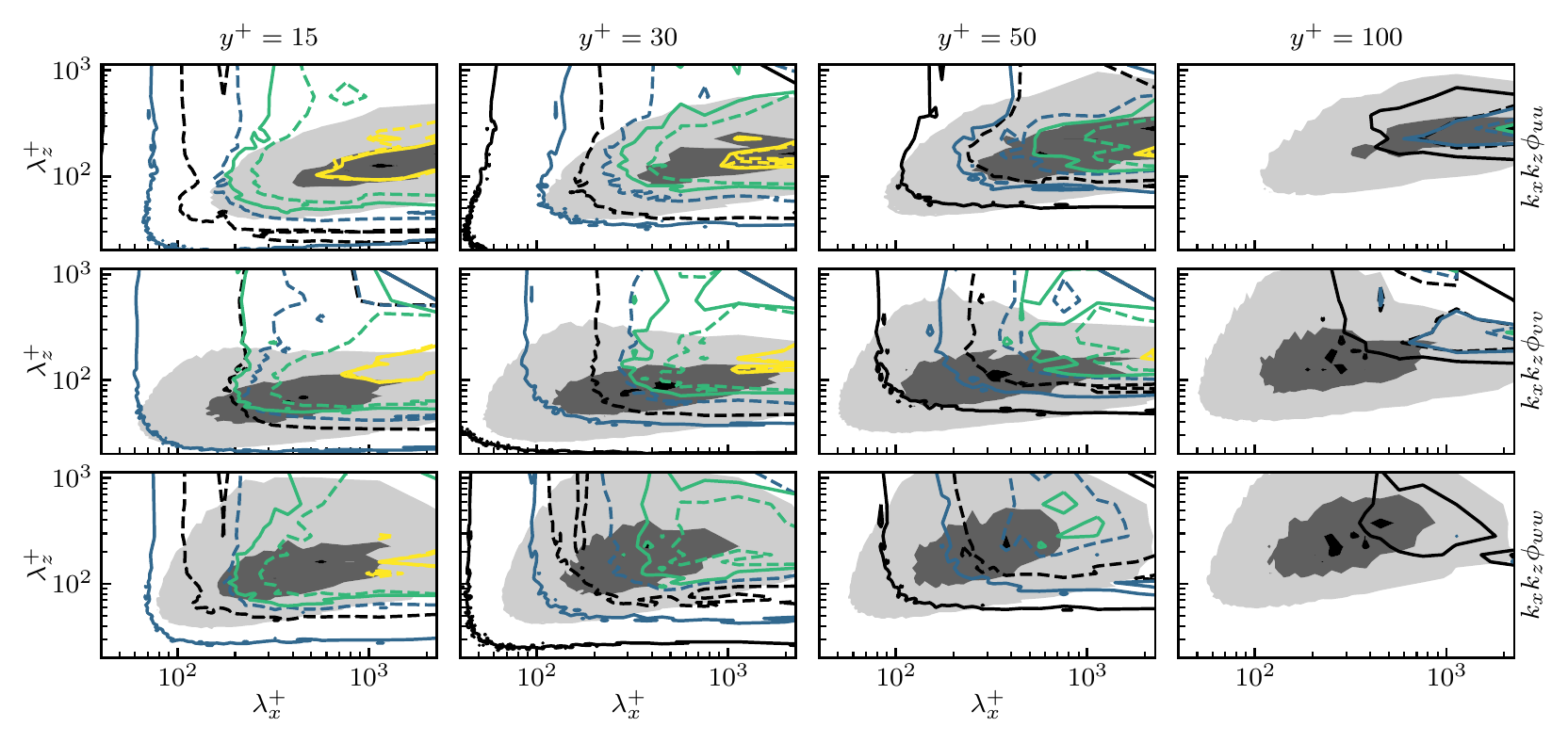}}
  \caption{Fractional spectral error for streamwise (first row), wall-normal (second row), and spanwise (third row) velocity fluctuations. From left to right, columns refer to inner-scaled wall-normal locations $y^+$ equal to 15, 30, 50, and 100. The contour level corresponds to $R_{ab}=0.5$. Contour lines refer to \lcap{-}{fd01} SRGAN, and \lcap{--}{fd01} FCN-POD predictions, respectively. Colours denote $f_d=1$ \sy{fd01}{s*}, $f_d=4$ \sy{fd04}{s*}, $f_d=8$ \sy{fd08}{s*}, and $f_d=16$ \sy{fd16}{s*}. Shaded contours refer to pre-multiplied two-dimensional power-spectral densities for the reference DNS data.}
\label{fig:07}
\end{figure*}

The pre-multiplied two-dimensional energy spectra of the flow quantities at the four wall-normal locations discussed above are shown in Figure~\ref{fig:07}. 
As reported in Ref.~\onlinecite{guastoni2020convolutional}, the amount of energy captured by the predictions decreases moving farther from the wall. 
Moreover, it is important to note that the FCN-POD method is able to recover scales larger than the subdomain size, although a discontinuity in the spectra can be observed at that wavelength, especially in the wall-normal and spanwise components. 
With respect to the effect of using SRGAN as a reconstruction method, the findings presented above are corroborated by the spectra. 
The predictions generated with SRGAN recover a wider range of energetic scales in both the streamwise and spanwise wavelengths for the three velocity fluctuations, even up to the case $f_d=8$, while for $f_d=16$ both methods have been shown to provide the same mean-squared error. 
Nonetheless, it is also important to mention that for $f_d=16$ at $y^+=100$ no energetic scales above the 10\% of the DNS reference has been recovered in the wall-normal and spanwise  spectra. 
This also occurs in the spanwise fluctuation spectra at $y^+=50$, but only for the predictions generated without SRGAN.

While Figure~\ref{fig:04} shows a large error for the extreme cases i.e., those reconstructions farther from the wall or with high $f_d$ values, a visual inspection of Figure~\ref{fig:05} reveals that even in these case the SRGAN predictions are able to capture the large-scale organization of the instantaneous flow field.
Therefore, the large observed errors can be ascribed to the attenuation of the velocity fluctuations, which makes it necessary to define a metric that evaluates the error based on the scale wavelengths.
Following Ref.~\onlinecite{encinar2019logarithmic}, a spectral fractional error can be defined as:
\begin{equation}
    R_{ab}(k_x,y,k_z) = \frac{\mathcal{R}e\langle(a-a^{\dagger})(b-b^{\dagger})^*\rangle(k_x,y,k_z)}{\mathcal{R}e\langle ab^*\rangle(k_x,y,k_z)},
    \label{eq:04}
\end{equation}
\noindent where $k_x$ and $k_z$ are the wave numbers in the streamwise and spanwise directions respectively, superscripts `*' and `$\dagger$' refer to complex conjugate and estimated quantities respectively, $\mathcal{R}e$ denotes real part, while $a$ and $b$ stand for either $u$, $v$, or $w$. 
Note that equation~(\ref{eq:04}) is related to the linear cohrence spectrum \cite{baars2016spectral,encinar2019logarithmic,tanarro2020effect}.
Figure~\ref{fig:07} shows the iso-contours of $R_{ab}=0.5$ for each case.
The reconstruction performance of $f_d=1$ and $f_d=4$ in the viscous region, which covers entirely the wavelengths above 10\% energy content of the streamwise fluctations, is particularly remarkable.
It can also be observed that the farther from the wall the fewer small scales are recovered in the reconstruction.
A similar behaviour is observed when increasing the downsampling factor $f_d$.

\begin{figure}
  \centerline{\includegraphics[width=\columnwidth]{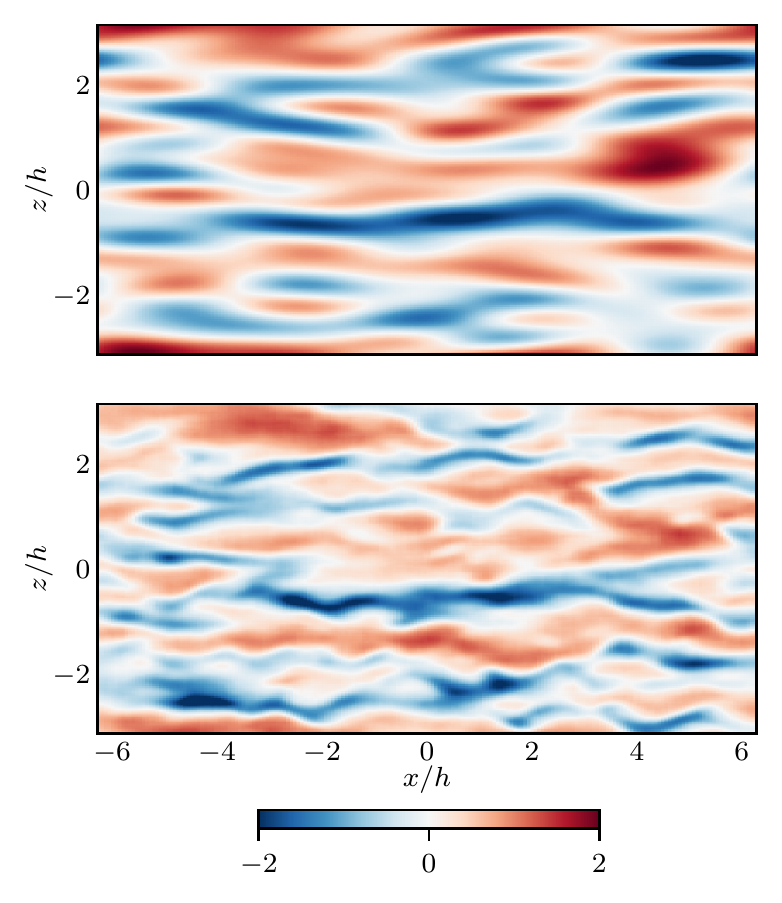}}
  \caption{Countour map of a sample streamwise velocity fluctuation field scaled with the corresponding standard deviation at $y^+=50$. Top and bottom panels denote reference filtered DNS and SRGAN prediction with $f_d=8$, respectively.}
\label{fig:08}
\end{figure}

Since the SRGAN predictions in the most challenging configurations exhibit some similarities with the filtered fields of the DNS reference, it is of interest to conduct the comparison.
Low-pass filtering has been applied to the DNS reference, where the cut-off lengths are adjusted to retain those scale with $R_{ab}<0.5$.
Figure~\ref{fig:08} shows this comparison for the case at $y^+=50$ with $f_d=8$ at the wall input data.
While the SRGAN prediction does not yield small-scale details, it exhibits a remarkable resemblance in terms of the streak patterns when compared with the filtered DNS.
Note that for the case of Figure~\ref{fig:08}, the cut-off wavelengths are set to $\lambda_x^+\approx500$ and $\lambda_z^+\approx100$.
If the mean-squared error displayed in Figure~\ref{fig:05} is computed with this filtered reference, the error reduces from 0.603 to 0.317 for the case of Figure~\ref{fig:08}.

\section{Summary and conclusions}\label{sec:concl}
The reconstruction of wall-parallel velocity fields from coarse measurements at the wall in a wall-bounded turbulent flow has been evaluated in this work, together with the resolution enhancement of the wall measurements. 
For that purpose, SRGAN has been proposed for both tasks. 
In the case of flow reconstruction from wall measurements, this architecture has been compared with the FCN-POD method proposed by Ref.~\onlinecite{guastoni2020convolutional}. 
The resolution enhancement of wall fields from their coarse counterparts has been carried out with donwsampling factors $f_d=[4,8,16]$. 
In the case of flow reconstruction, the methods have been evaluated at the following wall-normal locations: $y^+=[15,30,50,100]$ with wall downsampling factors $f_d=[1,4,8,16]$. 
SRGAN is shown to provide accurate reconstructions for the case of resolution enhancement of wall fields at $f_d=[4,8]$. 
In the most challenging case $(f_d=16)$, it can be observed that small-scale contributions are not recovered in the reconstruction, but the large-scale footprint of the flow at the wall is very well represented.
With respect to the flow reconstruction with full resolution at the wall, SRGAN is shown to provide a significant improvement with respect to the baseline FCN-POD method\cite{guastoni2020convolutional}.
The effect of increasing $f_d$ is also evaluated, showing a clear performance decrease unlike in the work of Ref.~\onlinecite{guemes2019sensing}, where only a weak effect is reported. 
This difference is ascribed to two reasons: First, the range of scales targeted in Ref.~\onlinecite{guemes2019sensing} only cover the large wavelengths, while this study does it for the entire spectrum.
Small-scale structures have characteristic time and length scales smaller than that of the filtering bandwidth from the coarse measurements, thus losing relevant information for the reconstruction.
Second, $f_d$ is not an adequate parameter to compare different databases of wall-bounded turbulent flows. 
To overcome this issue, we propose to use $\tilde{f}_{d}$, which takes into account the fraction of viscous length covered by a pixel. 
With this parameter, the effect of the downsampling is homogenized among the various works, showing a clear trend between the results of Ref.~\onlinecite{guemes2019sensing} and the ones presented here.
To the authors' knowledge this is the first study where DNNs are used to reconstruct flow fields from coarse wall measurements in a turbulent flow, and this approach has great potential in the context of closed-loop control.
In any case, it is observed that the accuracy improvements of SRGAN start to decrease when increasing $f_d$.
The capability of SRGAN to recover the large-scale structures present in the flow is also evaluated by means of the fractional spectral error\cite{encinar2019logarithmic} and assessment of filtered instantaneous fields.
It is shown that the SRGAN predictions are in good agreement with large-scale patterns obtained from the filtered DNS reference.

The present study has used high-resolution DNS data to train the proposed GAN network. 
However, the computational cost and requirements for producing this data increase with $Re_{\tau}$, being impossible to obtain for Reynolds numbers from real-life applications. 
Consequently, it is advisable to look for alternatives that allow the use of the proposed network in real-life scenarios. 
For instance, transfer learning could be explored as in Guastoni et al. (2021), to confirm that GANs are able to generalize from one $Re_{\tau}$ to another. 
Another option could be to rely on experimental data obtained from those real-life scenarios. 
However, experimental data might be contaminated by noise and have lower spatial and temporal resolution than well-resolved DNS data. 
Therefore, future investigations should focus on how to combine different neural networks with incomplete or noisy turbulent data.

\section*{Acknowledgements}
RV acknowledges the support by the G\"oran Gustafsson Foundation. 
SD and AI acknowledge the support by the European Research Council, under the COTURB grant ERC-2014.AdG-669505. 
HA acknowledges the support by Wallenberg AI, Autonomous Systems, and Software Program (WASP-AI).
We would also like to acknowledge Hampus Tober for useful discussions throughout the present study.

\section*{Data Availability Statement}
The data that support the findings of this study are available from the corresponding author upon reasonable request. 
The codes that support the findings of this study are openly available in GitHub at https://doi.org/10.5281/zenodo.5067426.

\bibliography{main}

\end{document}